\documentclass[11pt]{article}
\usepackage{color}
\usepackage[unicode,bookmarks,bookmarksopen,bookmarksopenlevel=2,colorlinks,linkcolor=blue,citecolor=green]{hyperref}
\usepackage{amsfonts}
\usepackage{amssymb,amsthm,amscd, amsbsy, array}
\usepackage{amsmath,bm}
\usepackage{bm}
\usepackage{amsmath, amssymb}
\usepackage{graphicx}
\usepackage{epsfig}
\usepackage{epsf}

\renewcommand{\vec}{\bm}

\def\lf{\left(}
\def\rg{\right)}

\def\beq{\begin{equation}}
\def\eeq{\end{equation}}
\def\beqa{\begin{eqnarray}}
\def\eeqa{\end{eqnarray}}
\def\nn{\nonumber}

\def\smallover#1/#2{\hbox{$\textstyle\frac{#1}{#2}$}}

\let\ssection=\section
\renewcommand{\section}{\setcounter{equation}{0}\ssection}

\begin{document}

\title{ Waves in the Skyrme--Faddeev model and integrable reductions}
\author{ L. Martina$^{1,2}$\thanks{%
e-mail: Luigi.Martina@le.infn.it}, M.V. Pavlov$^{3}$, S.A. Zykov$^{1,4}$ \\
$^{1 }$ Dipartimento di Matematica e Fisica "E. De Giorgi" \\
Universit\`a del Salento \\
$^{2}$ Sezione INFN di Lecce. Via Arnesano, CP. 193\\
I-73 100 LECCE (Italy) \\
Sector of Mathematical Physics, \\
Lebedev Physical Institute of Russian Academy of Sciences,\\
Moscow, Leninskij Prospeckt, 53; \\
$^4$ Institute of Metal Physics, Ural Branch of RAS, Ekaterinburg, Russia; }
\maketitle

\begin{abstract}
In the present article we show that the Skyrme--Faddeev model possesses
nonlinear wave solutions, which can be expressed in terms of elliptic
functions. The Whitham averaging method has been exploited in order to
describe slow deformation of periodic wave states, leading to a quasi-linear system. 
The reduction to general hydrodynamic systems have been considered and it is compared with 
other integrable reductions of the system.
\end{abstract}


\vspace{5mm} \noindent

\newpage


\section{Introduction}

The main aim of the present paper is to study periodic and multi-periodic
solutions of the so-called Skyrme--Faddeev model. In the recent years a
special interest was deserved by the 3D static nonlinear \emph{Skyrme -- Faddeev } $%
\sigma $-model for the field ${\vec{\phi}}:\mathbb{R}^{3}\rightarrow \mathbb{%
S}^{2}$, the total free energy of which is given by
\begin{equation}
S_{SF}=\int d^{3}x\left[ \frac{1}{4}\rho ^{2}\left( \partial _{k}{{\vec{\phi}%
}}\right) ^{2}+H_{ik}^{2}\right] ,\quad \left\{
\begin{array}{ll}
H_{ik}={\vec{\phi}}\cdot \left[ \partial _{i}{\vec{\phi}}\times \partial _{k}%
{\vec{\phi}}\right] , &  \\
{\vec{\phi}}\cdot {\vec{\phi}}=1, &
\end{array}%
\right. \label{Skyrme-Faddeev-Static}
\end{equation}%
where $\rho $ is a real positive constant and $H_{ik}$ is an antisymmetric
tensor field, the so-called Mermin - Ho vorticity \cite%
{Mermin,Pismen,Volovik} in Matter Physics, expressing the non irrotational
properties of the multi-component superfluid. On the other hand, after a
long work \cite{omat1,FaddeevNiemi}, it was proved that (\ref{Skyrme-Faddeev-Static}%
) can be seen as a special subcase, among many others \cite{Martone}, for
the background (classical) field of the quantum pure $SU\left( 2\right) $%
-Yang-Mills theory in the infrared limit, thus describing a self-consistent
\textquotedblleft mesonic\textquotedblright\ field in the context of the
Nuclear Physics.

The main interest is to look for localized finite energy solutions, for
which one imposes a constant value at spatial infinity, for instance $%
\lim_{\left\vert \mathbf{x}\right\vert \rightarrow \infty }{{\vec{\phi}}}%
=\left( 0,0,1\right) $, compactifying to $\mathbb{S}^{3}$ the space domain.
Then, from the homotopy group theory result $\pi _{3}\left( \mathbb{S}%
^{2}\right) ={\mathbb{Z}}$ \cite{Arnold}, one can conclude that all such
solutions (hopfions) are labelled by the so-called Hopf index $Q$ into
separated sectors. The Hopf index can be computed analytically, since $%
H_{ik} $ in the expression (\ref{Skyrme-Faddeev-Static}) is a closed 2-form on $%
\mathbb{S}^{3}$, so derivable from a 1-form $a_{k}=-\frac{\varepsilon _{ijk}%
}{4\pi }\int_{S^{3}}\frac{\left( x_{i}-x_{i}^{\prime }\right) H_{ij}\left(
x^{\prime }\right) }{\left[ \left( x_{m}-x_{m}^{\prime }\right) ^{2}\right]
^{3/2}}d^{3}x^{\prime }$. Thus one has \cite{Vakulenko,Kundu,Gladikowski-Hellmund}
\begin{equation*}
Q=\frac{1}{16\pi ^{2}}\int d^{3}x\,\varepsilon _{ikl}a_{i}\partial _{k}a_{l},
\end{equation*}%
expressing the linking number of the pre-images of two independent points on
the target $\mathbb{S}^{2}$, then $Q=0$ for spherical solutions, 1 for
toroidal shaped vortices. Only approximated analytical solutions are known
\cite{MMZ}. Many times tangled hopfions have been confirmed by numerical
studies \cite%
{Faddeev-Niemi,Battye,Hietarinta-Salo1,Hietarinta-Salo2,Sutcliffe}, which
have produced a comprehensive analysis of solitons with $1\leq Q\leq 16$,
proving the existence of local energy minima with knotted structure. On the
other hand, global analytical considerations \cite{Vakulenko,Kundu,Gladikowski-Hellmund,Ward}
have shown that the energy of such a knot is bounded from below by $%
S_{SF}\geq C\;\pi ^{2}\;\rho \;|Q|^{3/4}$, where $C$ is a  constant evaluated numerically
\cite{Sutcliffe}. The main consequence of the above bound is that hopfions
of higher topological charge can be broken only by adding an extra energy
contribution for their disentanglement. However, also space extended
structures were considered in \cite{Protogenov} and in \cite{Ward2}. In the
latter it was shown that periodic collection of localized objects may
condensate in order to form periodic structure in the space. However, as
pointed out in \cite{Ward2} the appearance of extended multi-sheeted
structures, similar in some extent to the stripes found in \cite{Protogenov}, may be energetically more favorable. Thus, the quest for periodic
(possibly exact) solutions for the Skyrme--Faddeev model becomes more
interesting. On the other hand, in a series of papers \cite{AlvarezCo}, it
was shown that, by adding certain differential constraints to the equation of
motion of the Lorentz invariant version of  (\ref{Skyrme-Faddeev-Static}),  one may obtain completely
integrable sub systems, with infinitely many local conservation laws. This
was another hint in the direction of integrable reductions. On the
experience of the hydrodynamic reductions we tried different approaches,
which lead us to identify 1) quasi-periodic solutions described in terms of
elliptic integrals, 2) the average motion of periodic waves accordingly to
Whitham's method,
3) constraints on the solutions expressed in terms of several Riemann invariants. The paper is organized into an Introduction and  three further Sections,
 concerning the previously listed topics and final considerations are
included in Conclusions.

\section{Periodic Solution}

First we consider the Skyrme--Faddeev model in the 4-dimensional
relativistic space time \cite{Kundu}, defined on the space-time $M=\left\{
\left( x^{0},\ldots ,x^{3}\right) \right\} $ endowed with the
(pseudo)-Riemannian metric $\text{diag}\left( g_{\mu }\right) =\left(
+,-,-,-\right) $, given by the Lagrangian density
\begin{equation}
\mathcal{L}=\frac{1}{32\pi ^{2}}\;\left( {\partial }_{\mu }{\vec{\phi}}\cdot
\partial ^{\mu }{\vec{\phi}}-\frac{\lambda }{4}\left( \partial _{\mu }{\vec{%
\phi}}\times \partial _{\nu }{\vec{\phi}}\right) \cdot \left( \partial ^{\mu
}{\vec{\phi}}\times \partial ^{\nu }{\vec{\phi}}\right) \right) -\kappa
\left( 1-{\vec{\phi}}\cdot {\vec{\phi}}\right) ,  \label{Lagrangian0}
\end{equation}%
where $\lambda =\frac{16}{\rho ^{2}}>0$ is the scaling parameter describing
the breaking of the conformal symmetry and $\kappa $ is a Lagrangian
multiplier implementing the constraint ${\vec{\phi}}\in \emph{S}^{2}$. The
model (\ref{Skyrme-Faddeev-Static}) is obtained by setting ${\partial }_{0}{%
\vec{\phi}}\equiv 0$.

It is well known that the geometric constraint ${\vec{\phi}}\cdot {\vec{\phi}%
}=1$ can be realized in several ways, but here it seems useful to introduce
the polar representation
\begin{equation}
{\vec{\phi}}=\left( \sin \widetilde{\Theta }\sin \widetilde{\Phi },\sin
\widetilde{\Theta }\cos \widetilde{\Phi },\cos \widetilde{\Theta }\right) ,
\label{polar}
\end{equation}%
where $\widetilde{\Theta}$ and $\widetilde{\Phi}$ are suitable function on the variables $\left(
x^{0},\ldots ,x^{3}\right) $ to be determined. The Lagrangian (\ref%
{Lagrangian0}) becomes
\begin{equation}
\mathcal{L}_{p}=\frac{1}{32\pi ^{2}}\;\left\{ \widetilde{\Theta }_{\mu }%
\widetilde{\Theta }^{\mu }+\sin ^{2}\widetilde{\Theta }\left[ \widetilde{%
\Phi }_{\mu }\widetilde{\Phi }^{\mu }-\frac{\lambda }{2}\left( \widetilde{%
\Theta }_{\mu }\widetilde{\Theta }^{\mu }\widetilde{\Phi }_{\nu }\widetilde{%
\Phi }^{\nu }-\widetilde{\Theta }_{\mu }\widetilde{\Theta }_{\nu }\widetilde{%
\Phi }^{\mu }\widetilde{\Phi }^{\nu }\right) \right] \right\} .  \label{i}
\end{equation}
Now, it is well known \cite{MMZ} that the equations of motion given by (\ref%
{Skyrme-Faddeev-Static}) admit harmonic plane wave solutions of the of the
form
\begin{equation}
{\vec{\phi}}=\left( \frac{2A\cos \left( p_{i}x^{i}\right) }{A^{2}+1},\frac{%
2A\sin \left( p_{i}x^{i}\right) }{A^{2}+1},\frac{1-A^{2}}{A^{2}+1}\right) ,
\label{linearwaves}
\end{equation}%
where $A$ parameterizes the amplitude of the third component and remarkably
the dispersion law is given by $p_{i}p^{i}=0$ . Actually, by using the
symmetry group of the model one can find a 12 parametric family (4
translations, 4 boosts/rotations and 3 gauge transformations) of solutions.
In particular, the axis of precession can be arbitrarily fixed by gauge
transformations. Nevertheless, they cannot be superimposed, because of the
nonlinearity character of the equations of motion. To have a visualization
of (\ref{linearwaves}) can think to a periodic assembly of vectors, whose
wave front are orthogonal to the direction ${\vec{p}}=\left(
p_{1},p_{2},p_{3}\right) $, and maintaining constant the projection $\phi
_{3}$. The configuration is similar to a cholesteric liquid crystal.
Moreover the energy density of the configuration is constant in the whole
space, being equal to $\mathcal{E}=\frac{A^{2}{\vec{p}}^{2}}{4\pi ^{2}\left(
1+A^{2}\right) ^{2}}$. Notice how solutions with smaller third component are
more energetic.

Looking for the simplest generalization of (\ref{linearwaves}), one assumes
that
\begin{equation}
\widetilde{\Theta }=\Theta \left[ \theta \right] ,\text{ }\widetilde{\Phi }%
=\Phi \left[ \theta \right] +\tilde{\theta},\text{ where \ }\theta =\alpha
_{\mu }x^{\mu },\tilde{\theta}=\beta _{\mu }x^{\mu }
\label{phase-pseudo-phase}
\end{equation}%
in which one distinguishes $\theta $ as the \textit{phase} from the \textit{%
pseudo-phase} $\tilde{\theta}$. From a different point of view, we are
looking for invariant solutions under a 8-parametric family of 2-dimensional
Abelian sub-algebra of the translation symmetry group given by
\begin{equation}
\left\{ {\vec{v}}_{\alpha }=\sum_{i=1}^{3}\alpha _{i}\mathbf{t}_{0}-\alpha
_{0}\sum_{i=1}^{3}\mathbf{t}_{i},{\vec{v}}_{\beta }=\sum_{i=1}^{3}\beta _{i}%
\mathbf{t}_{0}-\beta _{0}\sum_{i=1}^{3}\mathbf{t}_{i}\right\} ,
\label{salgebra}
\end{equation}%
where $\mathbf{t}_{i}$'s are the generators of the translations. Actually,
by using the adjoint action of the space-time rotational subgroup, we can
conjugate each of the above subalgebras to exactly one representative
sub-algebra of the form (\ref{salgebra}) belonging to a 3-parametric
sub-family. However, it is more easy to deal with all components for notational
homogeneity. Thus the equations of motion reduce to the announced
3-parametric family
\begin{eqnarray}
&&\left[ 2B_{3}-\frac{\lambda }{4}\mathcal{B}\sin ^{2}\Theta \right] \Theta
_{\theta \theta }=\sin 2\Theta \left( \frac{\lambda }{8}\mathcal{B}\;\Theta
_{\theta }^{2}+B_{3}\Phi _{\theta }^{2}+B_{2}\Phi _{\theta }+B_{1}\right) \\
&&2B_{3}\sin ^{2}\Theta \;\Phi _{\theta \theta }+\Theta _{\theta }\sin
2\Theta \left( 2B_{3}\Phi _{\theta }+B_{2}\right) =0,
\end{eqnarray}%
where $B_{1}=-\beta _{\mu }\beta ^{\mu }$, $B_{2}=-2\alpha _{\mu }\beta
^{\mu }$ $B_{3}=-\alpha _{\mu }\alpha ^{\mu }$ and $\mathcal{B}%
=B_{2}^{2}-4B_{1}B_{3}$.

Of course these equations provide the above linear solutions (\ref%
{linearwaves}), setting $B_3 = 0$ and $B_2 \neq 0$, then $\Theta = 2 \arctan
A$, $\Phi = \Phi_0 - \frac{B_1 }{B_2} \theta $ and $p_i = \frac{B_2 \beta_i
- B_1 \alpha_i}{B_2}$. On the other hand, for $B_3 \neq 0$, the solution is
given by $\Theta = 2 \arctan A$ and $\Phi = \Phi_0 -\frac{B_2\pm \sqrt
\mathcal{B} }{2 B_3} \theta $, with an analogous expression for the $p_i$'s.

To deal with the general situation one uses the expression of the
energy-stress tensor $T^{\mu \nu }=\left( \widetilde{\Theta }^{\nu }{%
\partial }_{\widetilde{\Theta }_{\mu }}+\widetilde{\Phi }^{\nu }{\partial }_{%
\widetilde{\Phi }_{\mu }}\right) \mathcal{L}_{P}-g^{\mu \nu }\mathcal{L}_{P}$%
. Substituting in it the ansatz (\ref{polar}), only derivatives with
respect to $\theta $ survive. Thus, the further vanishing divergence is
equivalent to take ${\partial }_{\theta }$ over a quantity obtained by
contracting $\mathcal{E}^{\mu }=T^{\mu \nu }\alpha _{\nu }$, corresponding
to total conserved quantities for the wave, seen as a function of the phase.
Their expressions are
\begin{eqnarray}
\mathcal{E}^{0} &=&\frac{-1}{32\pi ^{2}}\left\{ B_{3}\alpha _{0}\Theta
_{\theta }^{2}+\sin ^{2}\Theta \left[ 2{\vec{\alpha}}\cdot {\vec{\beta}}%
\,\beta _{0}+\left( B_{1}-2{\vec{\beta}}^{2}\right) \alpha _{0}\right.
\right.  \notag \\
&&\left. \left. +B_{3}\left( 2\beta _{0}+\alpha _{0}\Phi _{\theta }\right)
\Phi _{\theta }-\frac{\lambda \mathcal{B}}{8}\alpha _{0}\Theta _{\theta }^{2}%
\right] \right\} ,  \label{conslaw} \\
\ \mathcal{E}^{i} &=&\frac{-1}{32\pi ^{2}}\left\{ B_{3}\alpha _{i}\Theta
_{\theta }^{2}+\sin ^{2}\Theta \left[ B_{2}\beta _{i}-B_{1}\alpha
_{i}+B_{3}\left( 2\beta _{i}+\alpha _{i}\Phi _{\theta }\right) \Phi _{\theta
}-\frac{\lambda \mathcal{B}}{8}\alpha _{i}\Theta _{\theta }^{2}\right]
\right\} .  \notag
\end{eqnarray}%
These equations can be used to find an expression of $\Theta _{\theta }^{2}$
and $\Phi _{\theta }$. Precisely, assuming $B_{3}\neq 0$ one finds
\begin{eqnarray}
\Theta _{\theta }^{2} &=&\frac{8B_{3}\left( B_{1}\sin ^{2}\Theta
+U_{3}\right) -2B_{2}^{2}\left( \sin ^{2}\Theta +U_{2}^{2}\csc ^{2}\Theta
\right) }{B_{3}\left( 8B_{3}-\lambda \mathcal{B}\sin ^{2}\Theta \right) },
\label{Thetaeq} \\
\Phi _{\theta } &=&-\frac{B_{2}\left( U_{2}\csc ^{2}(\Theta )+1\right) }{%
2B_{3}},  \label{Phieq}
\end{eqnarray}%
where the $U_{i}$'s are two constants completely defining the quantities in (%
\ref{conslaw}) by the expressions $\mathcal{E}^{\mu }=U_{3}\alpha _{\mu }+%
\frac{B_{2}U_{2}}{2}\left( \frac{B_{2}}{B_{3}}\alpha _{\mu }-2\beta _{\mu
}\right) $. Thus, one has reduced the problem to the quadratures,
introducing only two new integration constants besides $\left\{
U_{2},U_{3}\right\} $, which determine the amplitudes of the phases. Finally
similar conclusions can be obtained in the case $B_{3}=0$. Despite of its
involved expression, equation (\ref{Thetaeq}) can be set in algebraic form
by the transformation
\begin{equation}
\Theta =\arcsin \sqrt{\psi },
\end{equation}%
forcing to be $0\leq \psi \leq 1$ and to be satisfied the equation
\begin{equation}
\psi _{\theta }^{2}=\frac{64(\psi -1)\left( \psi -A_{1}\right) \left( \psi
-A_{2}\right) }{\lambda ^{2}\mathcal{B}\psi _{1}\left( \psi _{1}-\psi
\right) },  \label{psieq}
\end{equation}%
where one has defined the constants $A_{1,2}=\frac{2B_{3}U_{3}\pm \sqrt{%
4B_{3}^{2}U_{3}^{2}-\mathcal{B}\, U_{2}^{2}}}{\mathcal{B}}$, related 1-1 to the values of the integrals of motion,   and  $\psi _{1}=%
\frac{8B_{3}}{\lambda \mathcal{B}}$. Assuming $A_{i}$ to be real and setting
$0<A_{1}<A_{2}<1$ for instance, by a continuous variation of $\psi _{1}$ one
obtains different behaviors of oscillation amplitudes for $\psi $: $A_{2}<\psi <1$ for
$\psi _{1}<0$, $A_{1}<\psi <A_{2}$ for $0<\psi _{1}<A_{2}$, $A_{1}<\psi <1$
for $\psi _{1}=A_{2}$, $A_{2}<\psi <1$ for $A_{2}<\psi _{1}$. So for all
choices of $\psi _{1}$ there is only one oscillating solution, bounded
between two of the three zeros of the numerator in (\ref{psieq}), even if
real unbounded solution may appear or complex ones (see Figure 1).

Analytically, equation (\ref{psieq}) can be integrated in terms of
incomplete elliptic integrals of the third kind. Precisely, by introducing a
parametric variable $Z$ one obtains the parametric form
\begin{eqnarray}
\theta \left( \psi \right) &=&\theta _{0}+\frac{1}{4}\sqrt{\frac{\mathcal{B}%
\lambda ^{2}\psi _{1}\left( \psi _{1}-A_{1}\right) ^{2}}{\left(
A_{1}-1\right) \left( A_{2}-\psi _{1}\right) }}\Pi \left[ \frac{A_{1}-A_{2}}{%
\psi _{1}-A_{2}};Z|\frac{(\psi _{1}-1)\left( A_{1}-A_{2}\right) }{\left(
A_{1}-1\right) \left( \psi _{1}-A_{2}\right) }\right] ,  \notag \\
\psi &=&-\frac{A_{2}\psi _{1}\sin ^{2}Z+A_{1}\left( \psi _{1}\cos
^{2}Z-A_{2}\right) }{A_{1}\sin ^{2}Z+A_{2}\cos ^{2}Z+\psi _{1}}
\end{eqnarray}%
which can be expressed in terms of Weierstrass $\mathcal{P}$ function. Furthermore,
from (\ref{Phieq}) the function $\Phi $ can be expressed again in terms of
incomplete elliptic integrals, namely
\begin{eqnarray}
\Phi &=&-\frac{B_{2}U_{2}}{2B_{3}}\left[ \int \frac{d\theta }{\psi \left(
\theta \right) }+\theta \right] +\Phi _{0}= \\
&-&\frac{s_{1}}{2\psi _{1}}\left[ \sqrt{\frac{2\psi _{1}\left( A_{1}-\psi
_{1}\right) {}^{2}\left( B_{1}\lambda \psi _{1}+2\right) }{\left(
A_{1}-1\right) \left( A_{2}-\psi _{1}\right) }}\Pi \left( \frac{A_{2}-A_{1}}{%
A_{2}-\psi _{1}};Z\left\vert \frac{\left( A_{1}-A_{2}\right) \left( \psi
_{1}-1\right) }{\left( A_{1}-1\right) \left( \psi _{1}-A_{2}\right) }\right.
\right) \right.  \notag \\
&&\left. +2s_{2}\sqrt{\frac{A_{2}\psi _{1}\left( A_{1}-\psi _{1}\right)
{}^{2}}{A_{1}\left( A_{1}-1\right) \left( A_{2}-\psi _{1}\right) }}\Pi
\left( \frac{\left( A_{1}-A_{2}\right) \psi _{1}}{A_{1}\left( \psi
_{1}-A_{2}\right) };Z\left\vert \frac{\left( A_{1}-A_{2}\right) \left( \psi
_{1}-1\right) }{\left( A_{1}-1\right) \left( \psi _{1}-A_{2}\right) }\right.
\right) \right] ,  \notag
\end{eqnarray}%
where $s_{1}=\text{sign\thinspace }B_{2}$, $s_{2}=\text{sign \thinspace }%
U_{2}$

In this parametric form it is evident that the phases of the spinorial field
and the phase $\theta $ do not have the same periodicity. Thus the solution
is generically quasi periodic and only for very special choices of the
parameters true periodic solutions appear (see  Figure 4 ).
Then it is convenient to adopt, as we will show in the next section, a
method which allows to describe solutions with a minimal set of parameters,
concerning only the periodicity in the phase. Here notice only that the
length-wave can be made very large when $A_{2}\rightarrow 1$ and $\psi
_{1}\rightarrow \infty $.

Then, an important observation occurs now: one has found a special
2-dimensional reduction of the Skyrme--Faddeev model, which is completely
integrable, and one may wonder if this is not in the class described in
\cite{AlvarezCo}. If one performs the transformation of the polar
representation of the field $\phi $ into the stereographic projection
\begin{equation}
w=i\tan \left( \frac{\Theta }{2}\right) \exp \left( -i \tilde{\Phi
}\right) ,w^{\ast }\rightarrow -i\tan \left( \frac{\Theta }{2}%
\right) \exp \left( i \tilde{\Phi}  \right) ,
\end{equation}%
the constraint imposed by the authors in \cite{AlvarezCo} is expressed by
\begin{equation}
{\partial }_{\mu }w\;{\partial }^{\mu }w=0.  \label{FerrConstr}
\end{equation}%
One easily verifies that it is satisfied by the harmonic wave solution (\ref%
{linearwaves}). On the other hand, if one replaces in (\ref{FerrConstr}) the
reduction in (\ref{phase-pseudo-phase}), depending on phase and
pseudo-phase, one obtains the relation
\beqa
& &\sin ^{2}(\Theta )\left( \Phi _{\theta }\left( B_{3}\Phi _{\theta
}+B_{2}\right) +B_{1}\right) -B_{3}\Theta _{\theta }^{2} +\nn \\ && i\, \sin (\Theta )\left( 2 B_{3} \Phi _{\theta }+B_{2}
\right)  \Theta _{\theta } =0
\eeqa
from which one sees that separately real and imaginary parts have to vanish.
Replacing the condition (\ref{Phieq}) into the last equation, one
obtains
\begin{equation}
-B_{3}\Theta _{\theta }^{2}+B_{1}\sin ^{2}(\Theta )-iB_{2}U_{2}\Theta
_{\theta }\csc (\Theta )-\frac{B_{2}^{2}\left( \sin ^{2}(\Theta
)-U_{2}^{2}\csc ^{2}(\Theta )\right) }{4B_{3}}=0,
\end{equation}%
saying that, excluding constant $\Theta $ solutions, or $B_{2}$ or $U_{2}$
have to be zero. But compatibility with equation (\ref{Thetaeq}) implies the
constancy of $\Theta $ in both cases. So, in conclusion, the solutions found
above are out of the sub-sector described by the constraint (\ref{FerrConstr}%
).

In the next Section we average the periodic solutions of the Skyrme--Faddeev
model by the Whitham approach.

\section{The Whitham averaging method}

The Whitham approach was developed for any multi-dimensional system
possessing  a Lagrangian formulation  (see details in \cite{Whitham}).
Nevertheless, only some year later, the first averaging  on a multi-dimensional example,  the well known three dimensional Kadomtsev-Petviashvili equation,  was provided by E. Infeld  \cite{Infeld}.  Now,  more
than thirty years later, we present the second multi-dimensional example, i.e. we are considering the  averaging of the four dimensional Skyrme--Faddeev
model. As we have seen above, this nonlinear system is determined by a Lagrangian (see (\ref{Lagrangian0})
and (\ref{i})) and possesses a multi-parametric family of periodic solutions (%
\ref{psieq}) (see also (\ref{Thetaeq}) and (\ref{Phieq})). In such a case,
following a  heuristic  approach, one can introduce
the averaged  Lagrangian $L(\gamma ,\omega ,\vec{\beta },\vec{k})$ 
in the new dynamical variables $\gamma ,\omega ,\vec{\beta },\vec{k}$, which  correspond to the derivatives with respect to space time variables of phase $\theta $ and
pseudo-phase $\tilde{\theta}$, now not necessarily linear as in (\ref{phase-pseudo-phase}). It means  that $\omega =-\theta _{x^0}, k^{i}=\theta
_{x^i}$ and $\gamma =-\tilde{\theta}_{x^0},\beta^{i}=\tilde{\theta}_{x^i}$.  Thus, one
immediately derives the four-dimensional quasilinear system%
\begin{equation}
\partial _{x^0}L_{\omega }=\partial _{x^i}L_{k^{i}},\text{ \ \ }\partial _{x^0}L_{\gamma
}=\partial _{x^i}L_{\beta ^{i}}, \end{equation}%
with the compatibility conditions
\beqa
k_{x^0}^{1}+\omega _{x^i}=0,\text{ \ \ },\text{ \ }k_{x^j}^{i}=k_{x^i}^{j} \text{ \ } i\neq j,  \label{s}
\\
\beta _{x^0}^{i}+\gamma _{x^i}=0,\text{ \ \ },\text{ \ }\beta _{x^j}^{i}=\beta _{x^i}^{j}  \text{ \ } i\neq j.\nn
\eeqa
The averaged Lagrangian density $L(\gamma ,\omega ,%
\vec{\beta },\vec{k})$ can be obtained in two steps. Formally, one replaces the family of periodic solutions (\ref{Thetaeq}) and (\ref{Phieq}) into the Lagrangian
\ref{i}, obtaining
\begin{equation*}
\hat{\mathcal{L}}_{p}=\sin ^{2}(\Theta )\left( -\frac{1}{2}\lambda \left( \frac{%
B_{2}^{2}}{4}-B_{1}B_{3}\right) \Theta _{\theta }^{2}+B_{3}\Phi _{\theta
}^{2}+B_{2}\Phi _{\theta }+B_{1}\right) +B_{3}\Theta _{\theta }^{2},
\end{equation*}%
which is a function only on $\theta$ thus, performing an integration over a finite space-like region, contributions from  $\theta$-independent  coordinates are just  time-independent finite multiplicative factors. Then, on a period of the wave, one leads to the Lagrangian
\begin{equation*}
L\equiv \frac{1}{2\pi }\oint \hat{ \mathcal{L}}_{p}\, d\theta ,
\end{equation*}
where, generalizing the standard Whitham approach, we need to introduce  two
natural normalizations (or constraints)
\begin{equation}
\oint d\theta  = 2\pi ,\text{ \ \ }<\Phi _{\theta }> = \oint  \Phi d\theta= 2 \pi m,  \label{norma}
\end{equation}%
where the integer \textquotedblleft $m$\textquotedblright\ is the number of
rotations  of the vector ${\vec{\phi}}$ around a  value determined by a given
pseudo-phase $\tilde{\theta}$. This situation is very similar to to the spin wave configurations called cyclon and extra-cyclon in multiferroic materials \cite{Lee}.   Then the corresponding
averaged Lagrangian density  is given by%
\begin{equation}
L  = \left( B_{1}-\frac{B_{2}^{2}}{4B_{3}}\right) \left( A_{1}+A_{2}+W\sqrt{%
\frac{\lambda }{2}B_{3}}\right) +\frac{B_{2}+2mB_{3}}{2B_{3}}\sqrt{%
A_{1}A_{2}(B_{2}^{2}-4B_{1}B_{3})},
\end{equation}%
where we introduced the function%
\begin{equation}
W=\frac{1}{2\pi }\oint \sqrt{\frac{\left( \psi -A_{1}\right) \left( \psi
-A_{2}\right) \left( \psi -\psi _{1}\right) }{1-\psi }}\frac{d\psi }{\psi }.
\end{equation}%
Thus,  one immediately can check that two equations $L_{A_{1}}=0$ and $%
L_{A_{2}}=0$ (see \cite{Whitham}) coincide with normalizations (\ref{norma}%
), while the Euler--Lagrange equations lead to four dimensional quasilinear system of the first order (\ref{s}).

\textbf{Remark}:  The two normalization conditions (\ref{norma}) formally allow
to exclude $A_{1}$ and $A_{2}$ from the above construction. This means that
solving (\ref{norma}), one can derive $A_{1}(\gamma ,\omega ,\mathbf{\beta },%
\mathbf{k})$ and $A_{2}(\gamma ,\omega ,\mathbf{\beta },\mathbf{k})$. Thus,
quasilinear system (\ref{s}) contains first order derivatives with respect
to $x^i$ of eight unknown functions ($\gamma ,\omega ,\mathbf{\beta },%
\mathbf{k}$) only. However, one can use the five roots $0,1,A_{1},A_{2}$ and $%
\psi _{1}$ for parametrization of periodic solution (\ref{psieq}).
Nevertheless, the function $W(\gamma ,\omega ,\mathbf{\beta },\mathbf{k})$
as well as the averaged Lagrangian cannot be expressed via complete elliptic
integrals of the first and second kind only. For this reason, all partial
derivatives $L_{\omega },L_{k^{i}},L_{\gamma },L_{\beta ^{i}}$ contain also
elliptic integrals of the third kind, thus  all expressions in quasilinear
system (\ref{s}) became too complicated to be presented explicitly in this
paper.

\section{The Method of Hydrodynamic Reductions}

In comparison with the approach considered in the previous Section, we
describe a special class of solutions of the Skyrme--Faddeev system, which
can be obtained by the method of hydrodynamic reductions (see \cite{Heavenly}).
The Lagrangian density $\mathcal{L}_{p}$  depends on $%
\widetilde{\Theta },\widetilde{\Theta }_{\mu }$ and $\widetilde{\Phi }_{\nu }
$. Actually,  the standard method of the hydrodynamic reductions is applicable to a
Lagrangian density  which depends on derivatives $\widetilde{\Theta }_{\mu }$
and $\widetilde{\Phi }_{\nu }$ only (see \cite{BFT}). Nevertheless, even in this
 more complicated case,  the method of hydrodynamic reductions can be
utilized. Indeed, the Euler--Lagrange equations%
\begin{equation*}
\partial _{\mu }\widetilde{\Theta }^{\mu }=\frac{1}{2}\sin (2\widetilde{%
\Theta })\tilde{\Phi}_{\nu }\tilde{\Phi}^{\nu }+\frac{\lambda }{2}\sin \widetilde{\Theta }%
\cdot \tilde{\Phi}_{\nu }\partial _{\mu }[\sin \widetilde{\Theta }(\widetilde{%
\Theta }^{\mu }\tilde{\Phi}^{\nu }-\widetilde{\Theta }^{\nu }\tilde{\Phi}^{\mu })],
\end{equation*}%
\begin{equation*}
\widetilde{\Theta }_{\mu }\tilde{\Phi}^{\mu }\sin (2\widetilde{\Theta })+\sin ^{2}%
\widetilde{\Theta }[\partial _{\mu }\tilde{\Phi}^{\mu }+\frac{\lambda }{2}%
\widetilde{\Theta }_{\nu }\partial _{\mu }(\tilde{\Phi}^{\mu }\widetilde{\Theta }%
^{\nu }-\tilde{\Phi}^{\nu }\widetilde{\Theta }^{\mu })]=0
\end{equation*}%
contain the field variable $\widetilde{\Theta }$, its first and second
derivatives $\widetilde{\Theta }_{\nu },\widetilde{\Theta }_{\mu \nu }$,
while the field variable $\tilde{\Phi}$ is involved only by its first and second
derivatives $\tilde{\Phi}_{\mu },\tilde{\Phi}_{\mu \nu }$. According to the method of
hydrodynamic reductions we introduce $N$ Riemann invariants $r^{i}(x,t,y,z)$, which satisfy simultaneously the three commuting diagonal hydrodynamic type systems%
\begin{equation*}
r_{x}^{i}=\mu ^{i}(\mathbf{r})r_{t}^{i},\text{ \ \ }r_{y}^{i}=\eta ^{i}(%
\mathbf{r})r_{t}^{i},\text{ \ \ }r_{z}^{i}=\zeta ^{i}(\mathbf{r})r_{t}^{i},
\end{equation*}%
where new auxiliary field variables $u^{\nu }\equiv \tilde{\Phi}^{\nu }$ and $%
w^{\mu }\equiv \widetilde{\Theta }^{\mu }$ depend on these Riemann
invariants only. For successful application of the hydrodynamic reduction
method, we should select all parts of these Euler--Lagrange equations
containing just first and second derivatives $\tilde{\Phi}_{\mu }, \tilde{\Phi}_{\mu \nu }$ and annihilating the coefficients of $\sin \widetilde{\Theta }$.   This yields the constraints%
\begin{equation}
\widetilde{\Theta }_{\mu }\tilde{\Phi}^{\mu }=0,\text{ \ \ }\tilde{\Phi}_{\nu }\tilde{\Phi}^{\nu
}\left( 1+\frac{\lambda }{2}\widetilde{\Theta }_{\mu }\widetilde{\Theta }%
^{\mu }\right) =0  \label{const}
\end{equation}%
implying the nonlinear system%
\begin{equation*}
\partial _{\mu }\widetilde{\Theta }^{\mu }=0,\text{ \ \ }\partial _{\mu
}\tilde{\Phi}^{\mu }+\frac{\lambda }{2}\widetilde{\Theta }_{\nu }\partial _{\mu
}(\tilde{\Phi}^{\mu }\widetilde{\Theta }^{\nu }-\tilde{\Phi}^{\nu }\widetilde{\Theta }%
^{\mu })=0,\text{ \ \ }\tilde{\Phi}_{\nu }\partial _{\mu }(\widetilde{\Theta }^{\mu
}\tilde{\Phi}^{\nu }-\widetilde{\Theta }^{\nu }\tilde{\Phi}^{\mu })=0.
\end{equation*}%
The latter is equivalent to the quasilinear system of the first order%
\begin{equation}
\partial _{\mu }w^{\mu }=0,\text{ \ \ }\partial _{\mu }u^{\mu } - \frac{%
\lambda }{2} w_{\eta }\partial _{\mu }(u^{\mu }w^{\eta}-u^{\eta }w^{\mu })=0,\qquad u_{\eta}\partial
_{\mu }(w^{\mu }u^{\eta }-w^{\eta }u^{\mu })=0,  \label{j}
\end{equation}
 so that  (\ref{const}) reads%
\begin{equation}
u_{\mu }w^{\mu }=0,\text{ \ \ } u_{\mu }u^{\mu }\left( 1 - \frac{\lambda }{2} w_{\beta }w^{\beta }\right) =0.
\end{equation}
The last relationship contains two admissible constraints%
\begin{equation}
u_{\mu }u^{\mu }=0,\text{ \ \ }1- \frac{\lambda }{2} w_{\beta }w^{\beta }=0.
\end{equation}%
Thus, the method of hydrodynamic reductions is applicable for (\ref{j}) if it is
equipped by the constraints%
\begin{equation}
u_{\mu }w^{\mu }=0,\text{ \ \ }u_{\mu }u^{\mu }=0 \label{constr1}
\end{equation}%
or by the two other constraints%
\begin{equation}
u_{\mu }w^{\mu }=0,\text{ \ \ }w_{\alpha }w^{\alpha}=\frac{2}{\lambda }. \label{constr2}
\end{equation}%
Actually, only   the second choice exists for the Skyrme--Faddeev model, while the
first one also is possible for $\lambda =0$, that is for the usual O(3) nonlinear $\sigma$-model.
Furthermore, let us notice the similarity of the previous constraints on the existence of hydrodynamical reductions for the Skyrme-Faddeev model with the constraint (\ref{FerrConstr}).

\section{Conclusions}
We have found exact analytic quasi-periodic spin waves for the Skyrme-Faddeev model.
 We have studied in detail their main features.  In particular they are determined in terms of elliptic integrals of third kind. Assuming that there exist solutions which are periodic, we have shown that the Lagrangian can be averaged by the Whitham method. This provides a Lagrangian for a set of parameters, describing the evolution of periodic waves in terms of a quasilinear system in partial derivatives of the first order.
Finally, noticing that in a general such a system like (%
\ref{s}) is non-integrable. But, one can use the so called
\textquotedblleft method of hydrodynamic reductions\textquotedblright,  which allows to extract particular
reductions, whose solutions can be parameterized by a set of arbitrary functions.  This can be achieved only if some special constraints (\ref{constr1}) or (\ref{constr2}) are satisfied. However, the analysis of those constraints deserve several technical complexities and it is not done here. Only, we remark that the constraints are a further restriction on the subsector of the solution space described in \cite{AlvarezCo}.

\subsection*{Acknowledgments}

The work was partially supported by the PRIN "Geometric Methods in the
Theory of Nonlinear Waves and their Applications " of the Italian MIUR and
by the INFN - Sezione of Lecce under the project LE41. The topic belongs to
the project \textquotedblleft Vortices, Topological Solitons and their
Excitations\textquotedblright\ in the frame of the agreement Consortium
EINSTEIN - Russian Foundation for Basic Researches. MVP's work was partially
supported by the RF Government grant \#2010-220-01-077, ag.
\#11.G34.31.0005, by the grant of Presidium of RAS \textquotedblleft
Fundamental Problems of Nonlinear Dynamics\textquotedblright\ and by the
RFBR grant 11-01-00197. M.V. Pavlov acknowledges the Department of
Matematica e Fisica "E. De Giorgi" for the warm hospitality. The authors acknowledge G. De Matteis for many helpful discussions.


\begin{figure}[]
\includegraphics[scale=1.2]{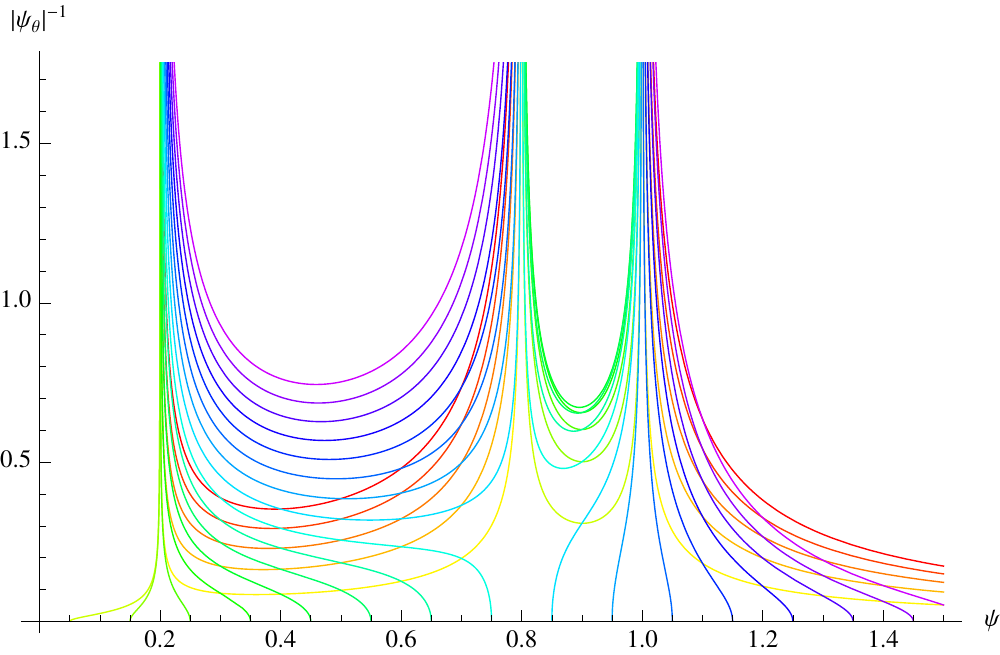}
\caption{The graphic for the inverse square root  of (\protect\ref%
{psieq})  for the  family of parameters  $\mathcal{B}=1, A_{1}=.1, A_{2}=.8$ and $-.45\leq \protect\psi %
_{1}\leq 1.55$ with steps of 0.1. Colors run accordingly from red to violet. Only one  bounded  periodic solutions  exists for any set of parameters.
The degenerate case $\protect\psi _{1}=A_{2}=.8$ is not considered, but it
corresponds to the confluence of the two yellow curves.}
\end{figure}
\begin{figure}[]
\includegraphics[width=14cm,height=2cm]{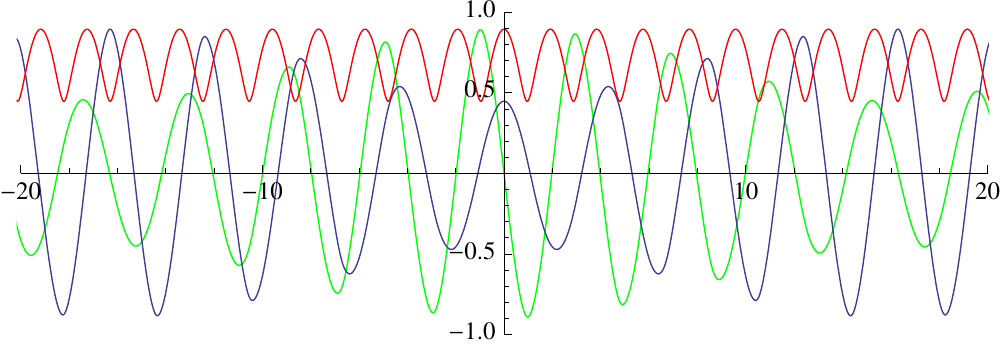}
\caption{The graphic for the $\protect\phi_1$ (green), $\protect\phi_2$
(blue) and $\protect\phi_3$ (red) as function of $x^3$ for a choice of the parameters $%
A_1=0.2,A_2=0.8,\protect\psi _1=0.9, \mathcal{B}=1,\protect\lambda %
=1,B_1=1,s_1=-1, s_2=-1$ . Accordingly, the wave vectors for the  phase and pseudo-phase have been chosen to be  $\alpha_\mu = \lf 0,0,0,0.33541\rg$ and $\beta_\mu= \lf 1.49638,1,0, -1.49638\rg $, respectively}
\end{figure}

\begin{figure}[]\includegraphics[width=14cm,height=2cm]{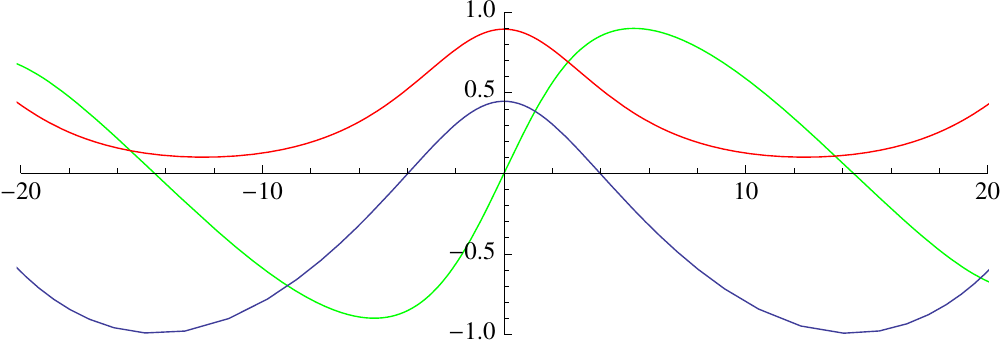}
\caption{The graphic for the $\protect\phi_1$ (green), $\protect\phi_2$(blue) and $\protect\phi_3$ (red) for a choice of the parameters $A_1=0.2,A_2=0.99,\protect\psi _1=20.01, \mathcal{B}=1,\protect\lambda =1,B_1=1,s_1=-1, s_2=-1$.  The wave vectors are $\alpha_\mu = \lf 0,0,0, -1.58153\rg$ and $\beta_\mu= \lf -1.04879,1,0, 1.04879\rg $, respectively }\end{figure}

\begin{figure}[]
\includegraphics[scale=1]{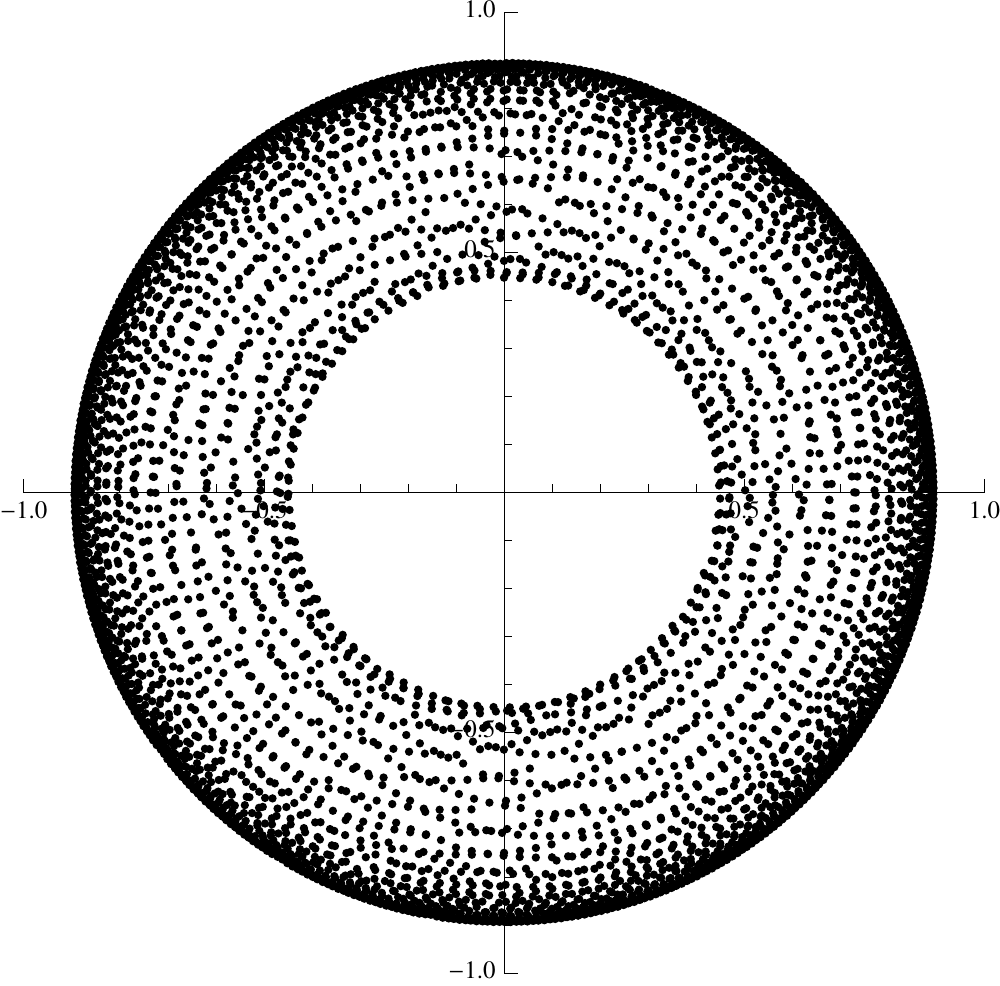}
\caption{ Projection on the plane $\left( \phi _{1},\phi _{2}\right) $ of a sample of about
5000 consecutive  values of the field $\phi $ along the axes $\lf 0, 0, x^3\rg$, for the same  choice of parameters as in Figure 2}
\end{figure}

\begin{figure}[]
\includegraphics[scale=1]{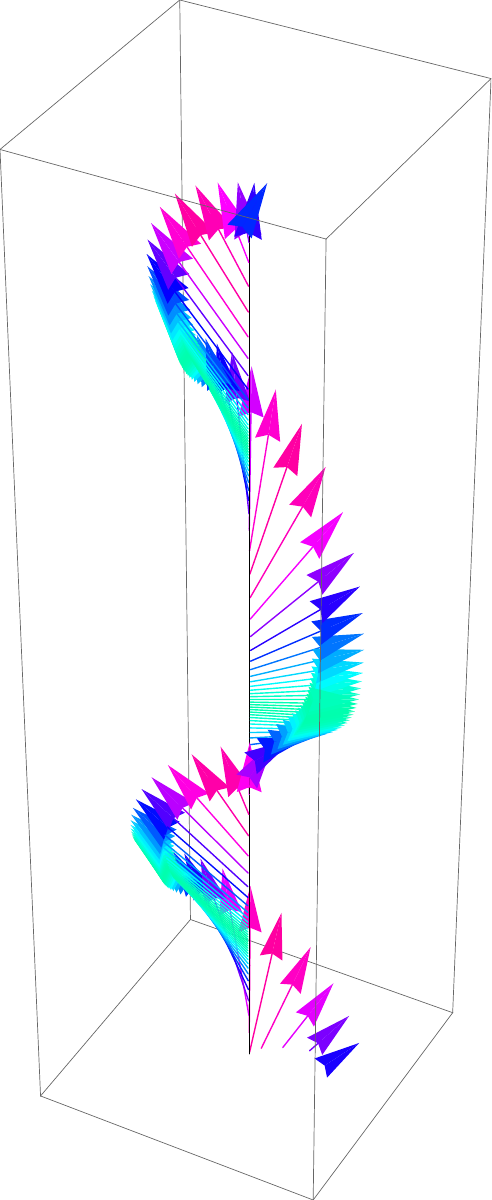}
\caption{A sample of 200 consecutive spin configurations along the $x^3$ axis, with parameters as in Figure 2.}
\end{figure}

\end{document}